\documentclass[
aps,%
12pt,%
final,%
notitlepage,%
oneside,%
onecolumn,%
nobibnotes,%
nofootinbib,%
superscriptaddress,%
noshowpacs]%
{revtex4}%
\usepackage{graphics}
\usepackage[T2A]{fontenc}
\usepackage[english]{babel}

\begin{document}

\title{Comments on the internal motion of massive charm quark in the process
of double charmonium production in  $e^+e^-$ annihilations}
\author{\firstname{A.~V.}~\surname{Berezhnoy}}
\email{aber@ttk.ru}
\affiliation{SINP of Moscow State University, Russia}

\begin{abstract}
The cross section value of 
doubly charmonium production in  $e^+e^-$ annihilations
have been estimated at interaction energy $\sqrt{s}=10.6$~GeV. 
The  quarkonium wave function shape, as well as the non-zero value 
of charm quark mass, have been taken into account.
\end{abstract}

\maketitle

\begin{figure}[!ht]
\vspace*{-2cm}
\centering
\resizebox*{1.05\textwidth}{!}{
\includegraphics*[150,350][550,900]{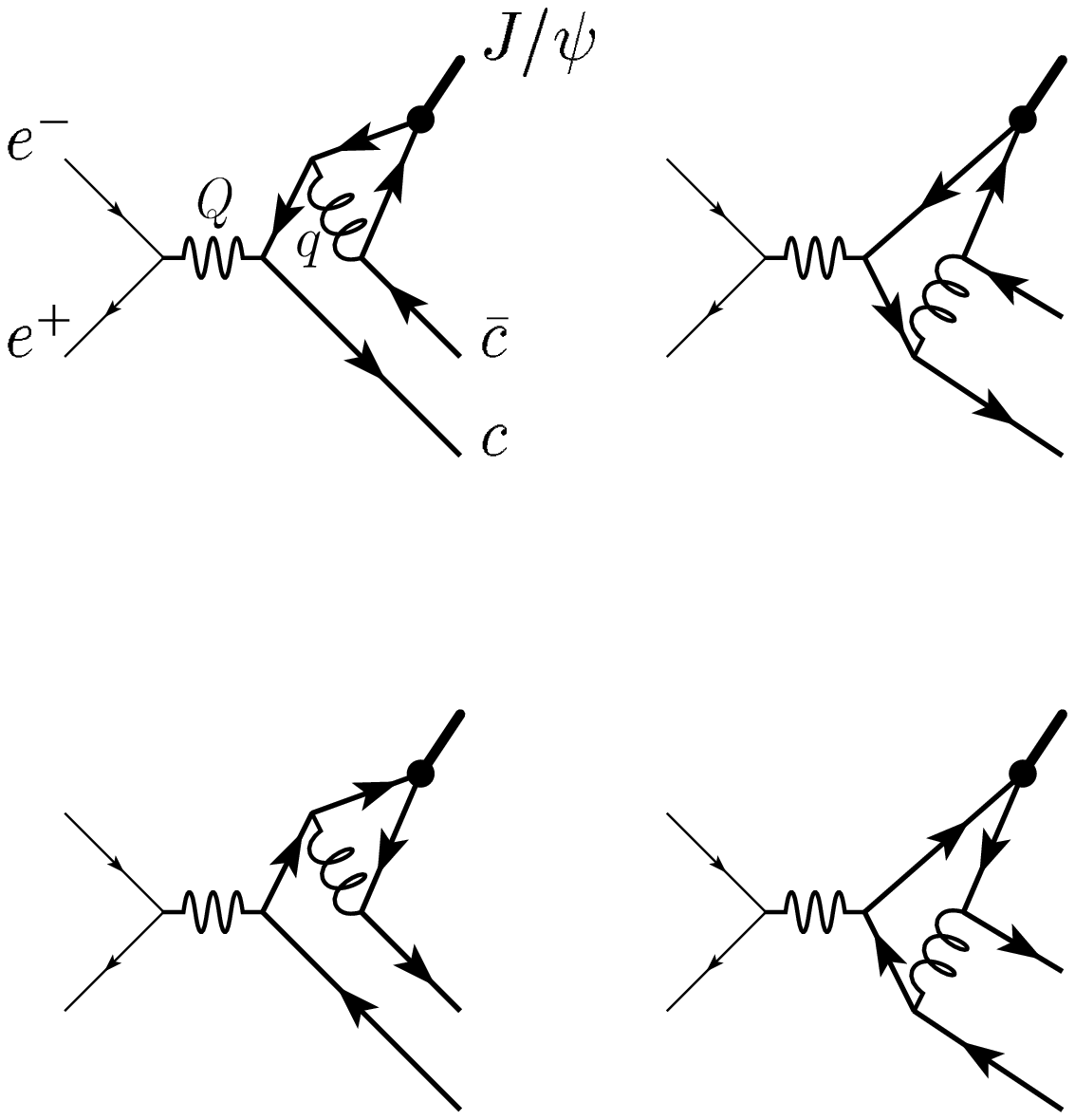}
\hfill
\includegraphics*[0,50][550,600]{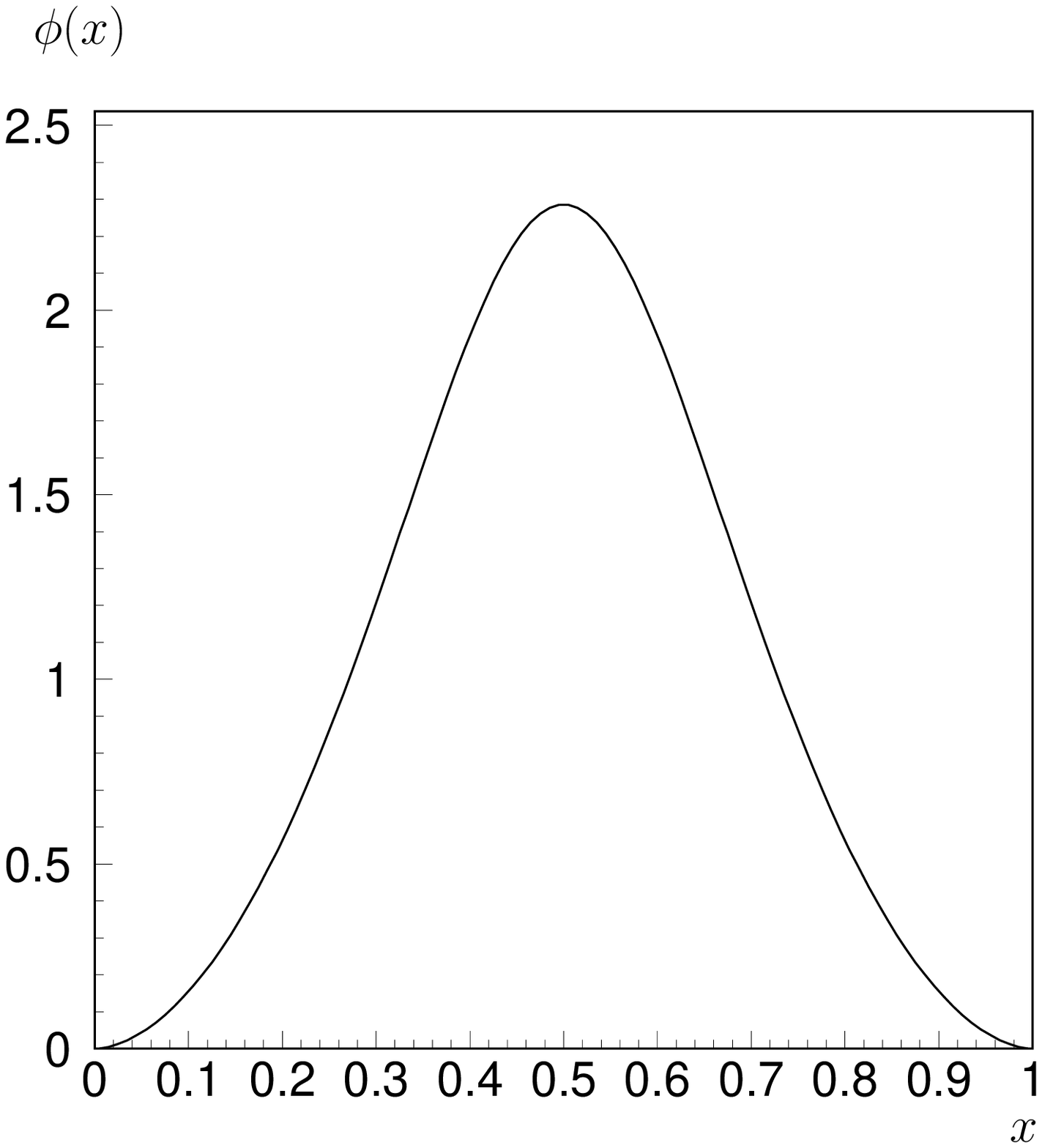}
}
\\
\parbox[t]{0.47\textwidth}{
\caption{ The leading order diagrams for the process $e^+e^-\to J/\psi +c \bar
c$. $Q$ is  momentum of the photon and 
$q$ is momentum of the intermediate quark or gluon.  The average  of $q^2$
in the total phase space is about several $m_c^2$.
If both $c\bar c$-pairs are fused into charmonia 
($e^-e^+\to J/\psi + \mbox{\rm charmonium}$), then $q^2\sim Q^2/4$ within the
$\delta$-approximation. 
\hfill}
\label{jpsicc_ccdual}}
\hfill
\parbox[t]{0.47\textwidth}{
\caption{The shape of charmonium wave function used 
in~\cite{lightcone_Bondar_Chernyak} within the lightcone
formalism (leading twist).}}
\end{figure}

The recent experimental results of the BELLE and BABAR 
Collaborations on  double charmonium production in $e^+e^-$ annihilations 
demand an essential revision
of the calculation techniques based on QCD factorization theorem.  
Indeed, the cross section value of charmonium pair production  
in $e^+e^-$ annihilations estimated within standard calculation
technique~\cite{2charmonium} underestimates the experimental 
data by an order of magnitude~\cite{Belle, BABAR}.
The detailed analysis has shown that one of the reasons of such underestimation 
is  large fixed virtualities of the intermediate quark and gluon,
$q^2\sim \frac{Q^2}{4}$ ($Q^2$ is a virtuality of the initial photon, 
see Fig.~\ref{jpsicc_ccdual}),
which occur in the standard calculations, where relative 
momenta of valence quarks of charmonium are not taken into account  in 
the hard part of amplitude (the so-called $\delta$-approximation).

Taking into account 
the relative motion of valence quarks allows to decrease contradictions
between theory and experiment. 
Calculations of the pair quarkonium production in $e^+e^-$ annihilations, 
 obtained  within a light cone formalism~\cite{lightcone_Bondar_Chernyak,
lightcone_Ma_Si, lightcone_Braguta_Likhoded_Luchinsky},
are in a qualitative agreement with the experimental data  of the 
BELLE and BABAR Collaborations. In these works, the $\delta$-shaped wave 
function of quarkonium
was replaced by a  function  $\phi(x)$, which is  ``spread'' on $x$, 
where $x$ is a momentum fraction of the quarkonium carried 
by a valence quark in the infinite momentum frame. 
Thus, the longitudinal ``internal motion'' of the valence quarks
is taken  into account (see also  \cite{KLP_frag}).  
This spreading  decreases the effective virtuality 
of the intermediate quark and gluon,  and, therefore,  
increases the predicted cross section value.

An analysis of doubly charmonium production 
within  quark-hadron duality also leads to the cross section value
increase~\cite{Berezhnoy_Likhoded_ccdual}, and allows to describe the data.

Recently several works have appeared, in which the internal motion 
effect has been estimated 
in the framework of NRQCD~\cite{NRQCD_Ebert_Martynenko,NRQCD_He_Fan_Chao}. 
The received $v^2$ 
corrections increase the cross section  by a factor about $1.5-2.5$. 
This value is less than value $\sim 10$ obtained within lightcone 
approximation~\cite{lightcone_Bondar_Chernyak,
lightcone_Ma_Si, lightcone_Braguta_Likhoded_Luchinsky}.  
Also there is a study~\cite{Bodwin_Kang_Lee}, where the critical analysis  
of lightcone formalism has been done for the double charmonium production in
$e^+e^-$ annihilations. The predictions obtained in this work are close to 
the NRQCD estimations~\cite{NRQCD_Ebert_Martynenko,NRQCD_He_Fan_Chao}.    

In this work we study the  dependence of internal motion 
effect from the wave function shape, as well as from the $c$ quark mass value. 
For this purpose we use a simple model described in~\cite{Brodsky_Lepage}  
instead of  rigorous lightcone formalism or NRQCD.  
Within the model under consideration, an amplitude of $J/\psi\eta_c$ production in
$e^+e^-$ annihilations is performed as follows: 
     
\begin{equation}
A_{J/\psi\eta_c} = \int\!\!\!\int dx\, dy\,\phi_{J/\psi}(x) \phi_{\eta_c}(y) T_{c\bar c c\bar c},
\label{convolution}
\end{equation} 
where $\phi_{J/\psi}(x)$ is a wave function of $J/\psi$, 
$x$ is $p_+$ fraction of $c$ quark inside $J/\psi$,
$\phi_{\eta_c}(y)$ is a wave function of $\eta$,  
 $y$ is $p_+$  fraction of $c$ quark inside $\eta_c$, and $T_{c\bar c c\bar c}$ 
is a hard part of amplitude,
which describes the process of four charm quarks production. The $c \bar c$-pairs
in $T_{c\bar c c\bar c}$ have quantum numbers corresponded to $J/\psi$ and
$\eta_c$ states. The transverse motion of quarks inside charmonium is neglected.

For $\phi_{J/\psi}$ and $\phi_{\eta_c}$ functions one choose the  
parametrization  used in work~\cite{lightcone_Bondar_Chernyak} 
within the lightcone 
formalism:
\begin{equation}
\phi_{lightcone}(x,v^2)=c(v^2)\phi^a(x)\left[\frac{x(1-x)}{1-4x(1-x)(1-v^2)}\right]^{1-v^2}, 
\label{lightcone_wave_function}
\end{equation} 
where $c(v^2)$ is a  normalization coefficient, $\phi^a(x)=6x(1-x)$ is an
asymptotic form of wave function 
and   $v$ is a typical velocity of quark-antiquark pair 
inside the charmonuim. In accord with the potential 
model for $J/\psi$ and $\eta_c$ mesons $v^2\approx 0.23$~\cite{potential_model}.

The results of these estimations have been compared with the predictions of
the $\delta$-approximation ($\phi(x,v^2)=\delta(x-1/2)$) and with the prediction
for the case of "tailless" wave function:

\begin{equation}
\phi_{tailless}(x,v^2)=
\left\{ 
\begin{array}{ll}
2.5 & \qquad 0.3<x<0.7 \\
0   & \qquad x<0.3,\;  x>0.7
\end{array}
\right.
\label{tailless_wave_function}
\end{equation} 

 The cross-section estimations within the model~(\ref{convolution}) for the 
 wave functions (\ref{lightcone_wave_function}) 
 and (\ref{tailless_wave_function}) have been done numerically.
  The standard expression for $\delta$-approximation
is 
\begin{equation}
\sigma \sim \frac{(1-4M^2/s)^{3/2}}{s^4},
\label{sigma_delta_approx}
\end{equation}
where $s$ is the $e^+e^-$ interaction energy squared, and $M^2$ is a charmonium
mass.

\begin{figure}[!ht]
\vspace*{-3.cm}
{\centering \resizebox*{0.8\textwidth}{!}
{\includegraphics{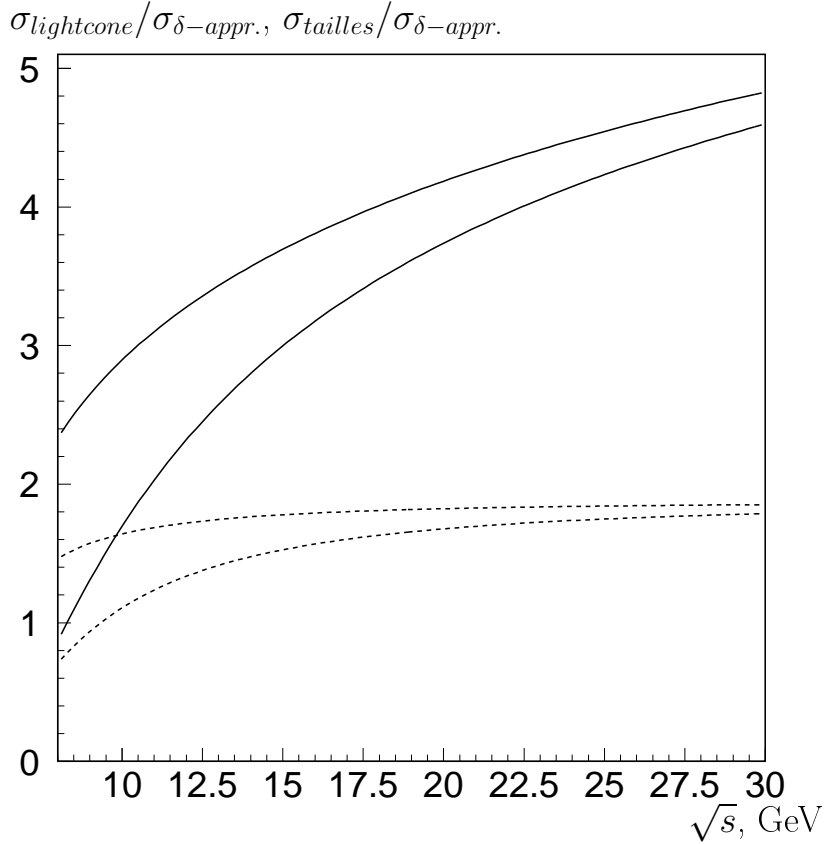}}
}
\caption{The ratio between the calculation results and
 the $\delta$-approximation
for different values of charm quark masses and wave function shapes
as a function of interaction energy:
upper solid curve is for 
$m_c=1.25$~GeV and the lightcone wave function~(\ref{lightcone_wave_function});
lower solid curve is for $m_c=1.5$~GeV and the lightcone wave 
function~(\ref{lightcone_wave_function}); 
upper dashed curve is for $m_c=1.25$~GeV and the ``tailless'' wave 
function~(\ref{tailless_wave_function});  
lower dashed curve is for $m_c=1.5$~GeV and the ``tailless'' wave 
function~(\ref{tailless_wave_function}). 
The $\delta$-approximation predictions are given for $m_c=1.5$~GeV. 
\hfill \label{ratio_to_delta}}
\end{figure}

The ratio between our calculation results and the $\delta$-approximation
are shown in Fig.~\ref{ratio_to_delta} for different values of charm quark
masses and wave function shapes. The $\delta$-approximation predictions are
given for $m_c=1.5$~GeV.

The toy model under discussion do not allows to shed light on  all aspects
of double charmonium production;
nevertheless the two main features of such process are clearly seen:
\begin{enumerate}
\item The cross section value is very sensitive to the $c$ quark mass value 
at interaction energies below 15~GeV.  
\item  The cross section value  depends strongly on the wave function behavior on
the tails (at $x<0.3$ and $x>0.7$).  
\end{enumerate}

As one can see from Fig.~\ref{ratio_to_delta}, the uncertainties caused by 
$c$ quark mass value are  large at energy 10.6~GeV, 
where Belle and BaBar data on doubly charmonium production have been obtained.
The cross section value for wave function (\ref{lightcone_wave_function})
is increased by factor 1.9 for $m_c=1.5$~GeV and by factor 3 
for  $m_c=1.25$~GeV in comparison to the $\delta$ approximation. 
For the "tailless" wave function (\ref{tailless_wave_function}) 
the cross section enhancement is about 20\% for $m_c=1.5$~GeV.
For  $m_c=1.25$~GeV the cross section value is increased by factor 1.6.
Unfortunately the tail behavior is not known with sufficient accuracy. 
Also it is not clear which $c$ quark mass should be used to obtain the
cross section. For the $\delta$-approximation, it should be better to use $m_c=M/2$
to obtain the correct meson mass. The situation is not so obvious if one takes
into account the quark internal motion, because this motion increases the
invariant mass of quark-aniquark system. This a reason to choose the
c quark mass value slightly smaller than $M/2$. It is worth to note, 
that  for the ``massless'' c quarks 
 the model under discussion leads to the cross-section increase by factor 10,
and, therefore, reproduce approximately the predictions of lightcone formalism.     

Contrary to the double charmonium production in $e^+e^-$ annihilation, 
the internal motion can be neglected for the process of associative charmonium
production $e^+e^-\to J/\psi c \bar c$. 
The cross section estimations for this process
have been done within the same model, as for the double charmonium production:
  \begin{equation}
A_{J/\psi c\bar c} = \int dx\,\phi_{J/\psi}(x) T_{c\bar c c\bar c}.
\end{equation}
 Our estimations  show 
 that the internal motion changes the cross section value by 5-10\% in
 comparison to $\delta$-approximation~\cite{Jpsi_cc_delta}. These results are
 in agreement with  predictions in~\cite{NRQCD_He_Fan_Chao}. The 
dependence on $c$ quark mass is more essential for this process.    
 
To summarize:
\begin{itemize}
\item Within the model under discussion the cross section value  
for the process $e^+e^-\to J/\psi\eta_c$ at 10.6 GeV can be increased by 
factor 1.2-3 depending on the c quark mass value and the wave function shape.
\item  In the frame of this model the dependence on the wave function shape
can be neglected for the process  $e^+e^-\to J/\psi c \bar c$.
\item The additional theoretical studies of charmonium production processes in
$e^+e^-$ annihilations are needed. 
\end{itemize}
 
Author thanks A.~K.~Likhoded and L.~K.~Gladilin for the fruitful discussion.
This work was partially supported by Russian Fund of Basic
Research (projects  No.~04-02-17530 and No.~05-07-90292),  
by the CRDF (project No.~M0-011-0), by President of Russian Federation 
(project for Young PhD Support No.~2773.2005.2), 
by Dynasty Foundation (project for Young Scientist Support), 
by INTAS Fellowship Grant for Young Scientists Nr.~05-112-5117,
and  was performed within the
program for Support of Leading Scientific Schools (project  No.~1303.2003.2).

\end{document}